\documentclass[12pt]{iopart}  \usepackage{graphicx}
\usepackage{subfigure,cite} \usepackage{amsfonts,amssymb}
\newcommand{\tinytext}[1]{\mbox{\tiny{#1}}}
\newcommand{\text}[1]{\mbox{\scriptsize{#1}}}

\begin{document}

\title[Polymer Translocation out of Planar Confinements]{Polymer
Translocation out of Planar Confinements}

\author{Debabrata Panja$^*$, Gerard T. Barkema$^{\dagger,\ddagger}$
and Robin C. Ball$^{**}$}

\address{$^*$Institute for Theoretical Physics, Universiteit van
Amsterdam, Valckenierstraat 65,\\ 1018 XE Amsterdam, The Netherlands

$\dagger$ Institute for Theoretical Physics, Universiteit Utrecht,
Leuvenlaan 4,\\  3584 CE Utrecht, The Netherlands

$^{\ddagger}$Instituut-Lorentz, Universiteit Leiden, Niels Bohrweg 2,
2333 CA Leiden,\\ The Netherlands

$^{**}$Department of Physics, University of Warwick, Coventry CV4 7AL,
UK}

\begin{abstract} 
Polymer translocation in three dimensions out of planar confinements
is studied in this paper. Three membranes are located at $z=-h$, $z=0$
and $z=h_1$. These membranes are impenetrable, except for the middle
one at $z=0$, which has a narrow pore. A polymer with length $N$ is
initially sandwiched between the membranes placed at $z=-h$ and $z=0$
and translocates through this pore. We consider strong confinement
(small $h$), where the polymer is essentially reduced to a
two-dimensional polymer, with a radius of gyration scaling as
$R^{\tinytext{(2D)}}_g \sim N^{\nu_{\tinytext{2D}}}$; here,
$\nu_{\tinytext{2D}}=0.75$ is the Flory exponent in two
dimensions. The polymer performs Rouse dynamics. Based on theoretical
analysis and high-precision simulation data, we show that in the
unbiased case $h=h_1$, the dwell-time $\tau_d$ scales as
$N^{2+\nu_{\tinytext{2D}}}$, in perfect agreement with our previously
published theoretical framework. For $h_1=\infty$, the situation is
equivalent to field-driven translocation in two dimensions. We show
that in this case $\tau_d$ scales as $N^{2\nu_{\tinytext{2D}}}$, in
agreement with several existing numerical results in the
literature. This result violates the earlier reported lower bound
$N^{1+\nu}$ for $\tau_d$ for field-driven translocation. We argue,
based on energy conservation, that the actual lower bound for $\tau_d$
is $N^{2\nu}$ and not $N^{1+\nu}$. Polymer translocation in such
theoretically motivated geometries thus resolves some of the most
fundamental issues that are the subjects of much heated debate in
recent times.
\end{abstract}


\maketitle

\section{Introduction\label{sec1}}

Polymer translocation through narrow pores in membranes is an active
field of research in recent times: as a cornerstone of many biological
processes, and also due to its relevance for practical applications.
Molecular transport through cell membranes is an essential mechanism
in living organisms. Often, the molecules are too long, and the pores
in the membranes too narrow, to allow the molecules to pass through as
a single unit. In such circumstances, the molecules have to deform
themselves in order to squeeze --- i.e., translocate --- themselves
through the pores. DNA, RNA and proteins are such naturally occurring
long molecules \cite{drei,henry,akimaru,goerlich,schatz} in a variety
of biological processes. Translocation is also used in gene therapy
\cite{szabo,hanss}, in delivery of drug molecules to their activation
sites \cite{tseng}, and as a potentially cheaper alternative for
single-molecule DNA or RNA sequencing \cite{expts1,nakane}.

\begin{figure}[h]
\begin{center}
\includegraphics[width=0.5\linewidth]{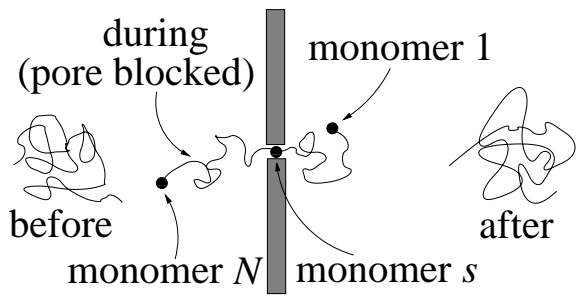}
\caption{Pictorial representation of a translocation event, with the
polymer shown before, during and after translocation. We number the
monomers starting with the end monomer on the side it moves to. The
number of the monomer located within the pore is $s$. \label{fig1}}
\end{center}
\end{figure}

In theoretical studies of translocation, the membrane is usually a
stationary object that does not show any movement or fluctuations, and
this also holds for the pore in it, through which translocation
occurs. The polymer is usually simplified to a sequentially connected
string of $N$ monomers. A central quantity in these theoretical
studies is the so-called dwell time $\tau_{d}$, which is the time the
pore remains blocked during a translocation event (Fig. \ref{fig1}).

The early theories of translocation were constructed in the spirit of
mean-field \cite{theory}, and that too for phantom polymers, wherein
translocation is quantified by a Fokker-Planck equation for
first-passage over an entropic barrier in terms of a single ``reaction
coordinate'' $s$. Here $s$ is the number of the monomer threaded at
the pore ($s=1,\ldots,N$), see Fig. \ref{fig1}. These mean-field type
theories apply under the assumption that every translocation step is
slower than the equilibration time-scale of the entire polymer.  Some
years ago, this assumption was questioned \cite{kantor1,kantor2},
wherein the authors provided lower bounds for $\tau_{d}$ for three
generic situations for phantom as well as self-avoiding polymers in
the absence of hydrodynamical interactions:
\begin{itemize}
\item[(a)] Unbiased translocation (i.e., translocation in the absence
of any driving field or force on the polymer), for which
$\tau_d\ge\tau_{\text{Rouse}}$, with $\tau_{\text{Rouse}}\sim
N^{1+2\nu}$ being the Rouse time, the longest time-scale in the
dynamics of the polymer;
\item[(b)] Translocation driven by a field $E$, acting on the polymer
only at the pore, for which it was shown that $\tau_d\ge N^{1+\nu}/E$;
\item[(c)] Translocation effected by a pulling force $F$ at the head
of the polymer, for which it was shown that $\tau_d\ge N^2/F$.
\end{itemize}
Here, $\nu$ is the Flory exponent: in three dimensions
$\nu\equiv\nu_{\text{3D}}\simeq0.588$ and in two dimensions
$\nu\equiv\nu_{\text{2D}}=0.75$. Accompanying numerical studies led
the authors to also suggest that the lower bounds indeed provide the
correct scalings for $\tau_d$; and based on these results, they
concluded that the dynamics of translocation, in (a-c), is anomalous
\cite{kantor1,kantor2}.

Subsequent numerical studies, however, did not immediately settle the
scaling for $\tau_d$ with $N$. In Tables \ref{table0a} and
\ref{table0b} we present a summary of results on the exponent for the
scaling of $\tau_d$ with $N$ (all results quoted are for self-avoiding
polymers in the absence of hydrodynamical interactions in the scaling
limit).
\begin{itemize}
\item[(i)] Unbiased translocation:
\begin{table}[!h]
\begin{center}
\begin{tabular}{c|c|c}
$\quad\mbox{authors}\quad$& $\quad\mbox{two-dimensions}\quad$&
$\quad\mbox{three-dimensions}\quad$\tabularnewline \hline Chuang {\it
et al.} \cite{kantor1}& $2.5$& $-$\tabularnewline \hline Luo {\it et
al.} \cite{luo1}& $2.50\pm0.01$& $-$\tabularnewline \hline Wei {\it et
al.} \cite{wei}& $2.51\pm0.03$& $2.2$\tabularnewline \hline Klein
Wolterink {\it et al.} \cite{wolt}& $-$& $2.40\pm0.05$\tabularnewline
\hline Milchev {\it et al.} \cite{milchev}& $-$&
$2.23\pm0.03$\tabularnewline \hline Dubbeldam {\it et al.}
\cite{dubbeldam}& $-$& $2.52\pm0.04$\tabularnewline \hline Panja {\it
et al.} \cite{anom}& $-$ & $2+\nu_{\tinytext{3D}}$ \tabularnewline
\hline This work & $2+\nu_{\tinytext{2D}}$ & $-$ \tabularnewline \hline
\end{tabular}
\caption{Existing results on the exponent for the scaling of $\tau_d$
with $N$ for unbiased translocation. \label{table0a}}
\end{center}
\end{table}
\vfill \eject
\item[(ii)] Field-driven translocation:
\begin{table}[!h]
\begin{center}
\begin{tabular}{c|c|c}
$\quad\mbox{authors}\quad$& $\quad\mbox{two-dimensions}\quad$&
$\quad\mbox{three-dimensions}\quad$\tabularnewline \hline Kantor {\it
et al.} \cite{kantor2}& $1.53\pm0.01$& $-$\tabularnewline \hline Luo
{\it et al.} \cite{luo2}& $1.72\pm0.06$ & $-$\tabularnewline \hline
Cacciuto {\it et al.} \cite{luijten}$^\dagger$& $1.55\pm0.04$ &
$-$\tabularnewline \hline Wei {\it et al.} \cite{wei}& $-$&
$1.27$\tabularnewline \hline  Milchev {\it et al.} \cite{milchev}&
$-$& $1.65\pm0.08$\tabularnewline \hline Dubbeldam {\it et al.}
\cite{dubbeldam2}& $-$& $1.5$\tabularnewline \hline This work & lower
bound $2\nu_{\tinytext{2D}}$; & $-$ \tabularnewline & observed $1.5$ &
\tabularnewline \hline
\end{tabular}
\caption{Existing numerical results on the exponent for the scaling of
$\tau_d$ with $N$ for field-driven translocation. The superscript
$\dagger$ indicates that we discuss this paper in detail in the
following paragraphs. \label{table0b}}
\end{center}
\end{table}
\end{itemize}

At a theoretical level, the lack of consensus on the scalings of
$\tau_d$ can easily be attributed to the fact that none of the works
in (i) and (ii) relates the numerically observed scaling of $\tau_d$
to the well-known dynamical features of polymers in a satisfactory
manner. We do note here that Refs. \cite{dubbeldam,dubbeldam2,klafter}
proposed to link the observed scalings of $\tau_d$ to anomalous
dynamics of translocation via a fractional Fokker-Planck equation; but
how this equation can be derived from the microscopic dynamics of a
single polymer, as well as the assumptions underlying the form of this
equation, remain entirely unclear.

In the recent past, this lack of consensus prompted us to investigate
the microscopic origin of the anomalous dynamics for unbiased
translocation. We set up a theoretical formalism, {\it based on the
microscopic dynamics of the polymer}, and showed that the anomalous
dynamics of translocation stem from the polymer's memory effects
\cite{anom,anomlong}, in the following manner. Translocation proceeds
via the exchange of monomers through the pore: imagine a situation
when a monomer from the left of the membrane translocates to the
right. This process increases the monomer density in the right
neighbourhood of the pore, and simultaneously reduces the monomer
density in the left neighbourhood of the pore.  The local enhancement
in the monomer density on the right of the pore \textit{takes a finite
time to dissipate away from the membrane along the backbone of the
polymer\/} (similarly for replenishing monomer density on the left
neighbourhood of the pore). The imbalance in the monomer densities
between the two local neighbourhoods of the pore during this time
implies that there is an enhanced chance of the translocated monomer
to return to the left of the membrane, thereby giving rise to
\textit{memory effects\/}. The ensuing analysis enabled us to provide
a proper microscopic theoretical basis for the anomalous dynamics,
leading us to conclude that $\tau_d$ scales as $\sim
N^{2+\nu_{\tinytext{3D}}}$ for unbiased polymer translocation in three
dimensions \cite{anom,anomlong}.

In Refs. \cite{anom,anomlong} we also showed that the theory presented
in Ref. \cite{dubbeldam} is not correct (which casts serious doubts
about the correctness of a related theory presented in the
theoretically related paper Ref. \cite{dubbeldam2}), but for unbiased
translocation in three dimensions, the numerical result $\tau_d\sim
N^{2.52\pm0.04}$ \cite{dubbeldam}, obtained by the use of a polymer
model very different from ours, is consistent with $\tau_d$ scaling as
$\sim N^{2+\nu_{\tinytext{3D}}}$. Two of us subsequently extended the
theoretical formalism Refs. \cite{anom,anomlong} to analyze
translocation by pulling the head of the polymer by a force $F$,
leading to the theoretical derivation for $\tau_d\sim N^2/F$
\cite{forced}.

The purpose of this paper is to push the theoretical formalism of
Refs. \cite{anom,anomlong,forced} further to study translocation in
three dimensions out of planar confinements for polymers performing
Rouse dynamics. Clearly, confinement reduces the number of
configurational states available to the polymer, reducing the
polymer's entropy and thereby increasing the polymer's free energy
\cite{casassa,degennes}. If the polymer is allowed to escape from the
confinement through a pore, then it will translocate out of the
confinement, and the free energy difference between the confined and
the free state of the polymer will drive translocation. While
confinement plays an important role for polymers in various biological
processes \cite{degennespap}, our interest in translocation out of
planar confinement in this paper stems more from a theoretical point
of view --- we aim to demonstrate that our theoretical formalism
\cite{anom,anomlong} works beautifully also in two dimensions. We
divide the three-dimensional space into two parts: $z>0$ and $z<0$ by
a membrane placed at $z=0$. This membrane is impenetrable to the
polymer except for a narrow pore. We then place two more parallel
completely impenetrable membranes at $z=-h$ and $z=h_1$. The polymer
is initially sandwiched between the membranes placed at $z=-h$ and
$z=0$. We only consider strong confinement of the polymer, i.e., $h\ll
R^{\tinytext{(3D)}}_g$, with the radius of gyration
$R^{\tinytext{(3D)}}_g$ for the polymer scaling in the present case as
$\sim N^{\nu_{\tinytext{3D}}}$. We study translocation out of planar
confinement for two separate cases: (1) $h_1=h$, and (2)
$h_1=\infty$. Our system for $h_1=\infty$ is shown below in
Fig. \ref{fig2}.
\begin{figure}[h]
\begin{center}
\includegraphics[width=0.5\linewidth]{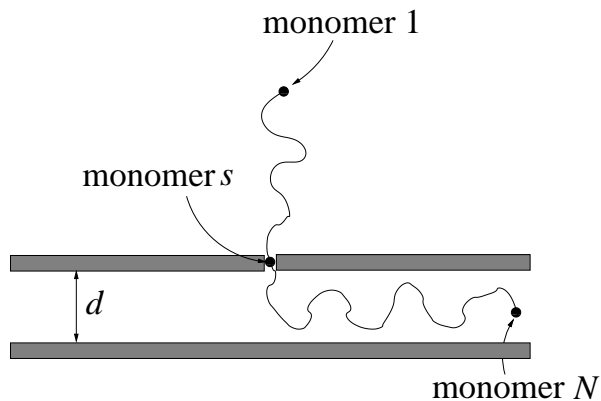}
\caption{Our system, and a snapshot of a translocating polymer out of
planar confinement, with $h_1=\infty$. We use strong confinement, for
which $h\ll R^{\tinytext{(3D)}}_g$, where $R^{\tinytext{(3D)}}_g$ is
the radius of gyration for the polymer, scaling as $\sim
N^{\nu_{\tinytext{3D}}}$.\label{fig2}}
\end{center}
\end{figure}

We substantiate our theoretical analysis with extensive Monte Carlo
simulations, in which the polymer performs single-monomer moves. The
definition of time is such that single-monomer moves along the
polymer's contour are attempted at a fixed rate of unity, while moves
that change the polymer's contour are attempted ten times less often.
Details of our self-avoiding polymer model in 3D can be found in
Refs. \cite{heukelum03,anomlong}.

\begin{figure}[ht]
\begin{center}
\includegraphics[width=0.5\linewidth]{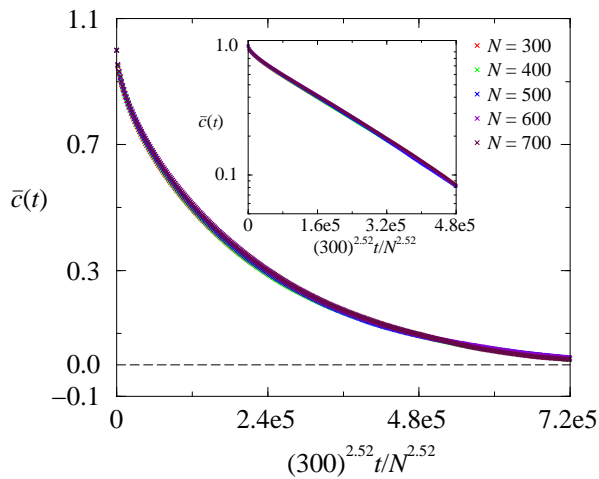}
\caption{End-to-end equilibrium correlation function $\bar{c}(t)$ for
a polymer with one end tethered at the pore on the membrane at $z=0$,
for $h=3$. The data for each $N$ are averaged over $256$
realisations. \label{fig3}}
\end{center}
\end{figure}

At strong confinements, i.e., with $h\ll N^{\nu_{\tinytext{3D}}}$ the
confined segment of the polymer essentially behaves as a
two-dimensional polymer. We demonstrate this below for $h=3$ (lattice
units). We tether one end of the polymer of length $N/2$ at the pore
on the membrane at $z=0$, confine the polymer between the plates at
$z=0$ and $z=-h$, and measure the end-to-end distance $R_e$ of the
polymer in equilibrium, as well as the equilibrium correlation
function of the end-to-end vector. That $R_e$ scales as
$(N/2)^{\nu_{\tinytext{2D}}}$ and the equilibrium correlation function
of the end-to-end vector behaves as $\exp[-t/\tau_{\text{Rouse(2D)}}]$
are demonstrated below in Table \ref{table1} and Fig. \ref{fig3}
respectively, with $\tau_{\text{Rouse(2D)}}$ being the Rouse time in
two dimensions, scaling, for a polymer of length $N/2$, as
$(N/2)^{1+2\nu_{\tinytext{2D}}}$.
\begin{table}[!h]
\begin{center}
\begin{tabular}{p{2cm}|p{2cm}|p{3cm}}
\hspace{6mm}$N/2$ &
\hspace{7mm}$\langle R_e\rangle$ &
\hspace{2mm}$\langle R_e\rangle/(N/2)^{\nu_{\tinytext{2D}}}$
\tabularnewline \hline \hline
\hspace{7mm}100 &
\hspace{5mm}10.248 &
\hspace{6mm}\quad0.324 \tabularnewline \hline
\hspace{7mm}150 &
\hspace{5mm}13.759 &
\hspace{6mm}\quad0.321 \tabularnewline \hline
\hspace{7mm}200 &
\hspace{5mm}17.135 &
\hspace{6mm}\quad0.322 \tabularnewline \hline
\hspace{7mm}250 &
\hspace{5mm}20.242 &
\hspace{6mm}\quad0.322 \tabularnewline \hline
\hspace{7mm}300 &
\hspace{5mm}23.171 &
\hspace{6mm}\quad0.321\tabularnewline \hline
\hspace{7mm}350 &
\hspace{5mm}26.213 &
\hspace{6mm}\quad0.324\tabularnewline \hline
\hspace{7mm}400 &
\hspace{5mm}28.872 &
\hspace{6mm}\quad0.323\tabularnewline \hline
\end{tabular}
\caption{Average end-to-end distance for a polymer, confined between
the membranes at $z=0$ and $z=-h$, with one end tethered at the pore
on the membrane at $z=0$, for $h=3$. The angular brackets denote an
average over 10,000 runs for each $N$.\label{table1}}
\end{center}
\end{table}

The confinement of a polymer of length $N$ between the two planes at
$z=0$ and $z=-h$ is accompanied by an entropic (or free energy) cost
of $\Delta F\sim Nh^{-1/\nu_{\tinytext{3D}}}$ \cite{degennes}. Thus,
in the case that $h_1>h$, the initial state of the polymer is
entropically less favourable, and the polymer will escape through the
pore to the wider space between the membranes at $z=0$ and
$z=h_1$. This process is analogous to field-driven translocation, with
a field strength $\sim
[h^{-1/\nu_{\tinytext{3D}}}-h_1^{-1/\nu_{\tinytext{3D}}}]$.  Although
this analogue has been correctly identified in Ref.  \cite{luijten},
the interpretation of $\tau_d$ therein is not correct:
Ref. \cite{luijten} used $\nu_{\tinytext{3D}}$ to describe the size of
the confined polymer, while Table \ref{table1} and Fig. \ref{fig3}
clearly show that the scaling of the size of the confined polymer with
length is characterised by the exponent $\nu_{\tinytext{2D}}$.

The proper interpretation of the numerical result of
Ref. \cite{luijten}, therefore, clearly violates the lower bound
$N^{1+\nu_{\tinytext{2D}}}$, provided by Ref. \cite{kantor2} for
$\tau_d$. Table \ref{table0b} shows more independent numerical
evidence that the scaling of $\tau_d$ for field-driven translocation
falls far short of $N^{1+\nu_{\tinytext{2D}}}$. This raises serious
doubts about the theoretical lower bound $N^{1+\nu}$ for $\tau_d$ for
field-driven translocation argued in Ref. \cite{kantor2}.

Our main results in this paper are two-fold. First, for $h_1=h$, the
entropic drive is absent; the translocation dynamics reduces to that
of an unbiased translocation for a two-dimensional polymer. Based on
theoretical analysis and high-precision simulation data, we show that
the dwell time for a polymer of length $N$ scales as $\tau_d\sim
N^{2+\nu_{\tinytext{2D}}}$ for $h_1=h$, in perfect agreement with our
previous results \cite{anom,anomlong}. In this paper, we actually go
one step further than Refs. \cite{anom,anomlong} to show that the
probability distribution of the dwell time, $P(\tau_d)$ for $h_1=h$
has a scaling form $P(\tau_d)\sim {\cal
P}(\tau_d/N^{2+\nu_{\tinytext{2D}}})/N^{2+\nu_{\tinytext{2D}}}$. Secondly,
for field-driven translocation, we argue, based on conservation of
energy, that the lower bound for $\tau_d$ for field-driven
translocation is given by $N^{2\nu}$ in the absence of
hydrodynamics. Using the analogue between translocation out of planar
confinement for $h_1=\infty$ and field-driven translocation in two
dimensions, we demonstrate numerically that $\tau_d\sim
N^{2\nu_{\tinytext{2D}}}$, with our numerical results being consistent
with those of Refs. \cite{kantor2} and \cite{luijten}. Study of
polymer translocation in these (theoretically motivated) geometries,
therefore, is a fine test case for the fundamental physics governing
translocation dynamics.

This paper is organised as follows. In Sec. \ref{sec2} we discuss a
method to measure $\Phi(t)$, the component of the polymer chain
tension at the pore, perpendicular to the membrane. In Sec. \ref{sec3}
we analyze the memory effects in $\phi(t)$, the imbalance of the
polymer chain tension across the pore. In Sec. \ref{sec4} we discuss
the consequence of these memory effects on unbiased translocation,
i.e., for the case $h_1=h$. In Sec. \ref{sec5} we derive the lower
bound $N^{2\nu}$ for $\tau_d$ for field-driven translocation, and
discuss the consequence of this lower bound and the memory effects on
translocation out of confinement for the case $h_1=\infty$. We finally
end this paper with a discussion in Sec. \ref{sec6}.

\section{Chain tension at the pore perpendicular to the
  membrane\label{sec2}}

A translocating polymer can be thought of as two segments of polymers
threaded at the pore, while the segments are able to exchange monomers
between them through the pore. In Ref. \cite{anom} we developed a
theoretical method to relate the dynamics of translocation to the
imbalance of chain tension between these two segments across the
pore. The key idea behind this method is that the exchange of monomers
across the pore responds to the imbalance of chain tension $\phi(t)$;
in its turn, $\phi(t)$ adjusts to $v(t)$, the transport velocity of
monomers across the pore. Here, $v(t)=\dot{s}(t)$ is the rate of
exchange of monomers from one side to the other, where $[s(t)-s(0)]$
is the total number of monomers translocated from one side of the pore
to the other in time $[0,t]$. In fact, we noted that $[s(t)-s(0)]$ and
$\phi(t)$ are conjugate variables in the thermodynamic sense, with
$\phi(t)$ playing the role of the chemical potential difference across
the pore.
\begin{figure}[!h]
\begin{center}
\includegraphics[width=0.6\linewidth]{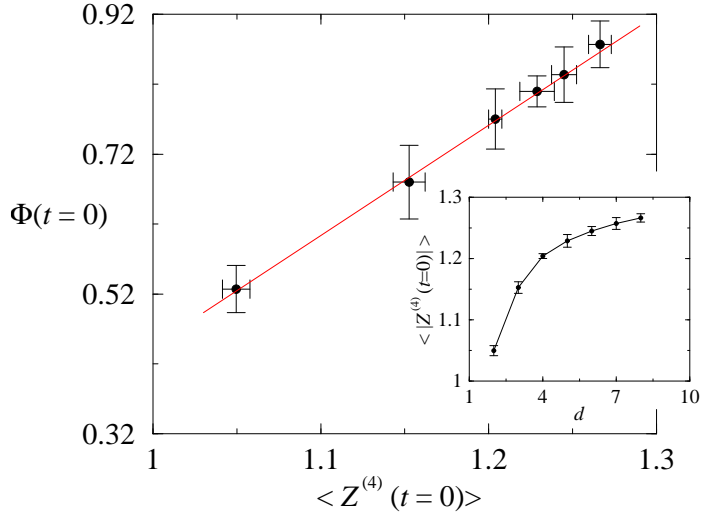}
\caption{$\langle Z^{(4)}(t=0)\rangle$ vs. $\Phi(t=0)$ demonstrating
the linear relationship between the two, for $N=100$ and $h=2,3,4,5,6$
and $8$ respectively. The angular brackets for $\langle
Z^{(4)}(t=0)\rangle$ indicate an average over $2,400$ polymer
realisations. The data for $\Phi(t=0)$ are obtained $2,400$ polymer
realisations as well. The solid line corresponds to the linear
best-fit. Inset: $\langle Z^{(4)}(t=0)\rangle$ as a function of
$h$.\label{fig4}}
\end{center}
\end{figure}

By definition, $\phi(t)=\Phi_{z>0}(t)-\Phi_{z<0}(t)$ where
$\Phi_{z>0}(t)$ and $\Phi_{z<0}(t)$ are respectively the chain tension
(or the chemical potential) on the $z>0$ and the $z<0$ side of the
pore. Consider a separate problem, where we tether one end of a
polymer to a fixed membrane, yet the number of monomers are allowed to
spontaneously enter or leave the tethered end, then we have
\begin{eqnarray}
\frac{W_t(-\rightarrow +)}{W_t(+\rightarrow-)}=\exp[\Phi(t)/k_BT]\,,
\label{e1}
\end{eqnarray}
where $W_t(-\rightarrow +)$ [resp. $W_t(+\rightarrow-)$] is the rate
that a monomer enters (resp. leaves) the polymer chain through the
tethered end at time $t$. Note that tethering the polymer while
allowing monomers to enter or leave the polymer at the tethered end is
precisely the case that translocation represents.

For a translocating polymer out of confinement between the membranes
at $z=0$ and $z=-h$, note that at $t=0$, it is easy to use
Eq. (\ref{e1}) to measure the chain tension for both segments at the
pore [$\Phi(t=0)$ in our notation], since under these conditions, we
also have the relation that
\begin{eqnarray}
P_-\,W_{t=0}(-\rightarrow +)=P_+\,W_{t=0}(+\rightarrow-)\,,
\label{e2}
\end{eqnarray}
where $P_-$ (resp. $P_+$) is the probability that the $z<0$ (or the
$z>0$) polymer segment has one monomer less (resp. one extra
monomer). Equations (\ref{e1}) and (\ref{e2}) together yield us
\begin{eqnarray}
\Phi(t=0)=k_BT\,\ln\frac{P_+}{P_-}\,.
\label{e3}
\end{eqnarray}

The chain tension as obtained from Eq. (\ref{e3}) is linearly related
to the distance of the centre-of-mass of the first few monomers along
the polymer's backbone, at the immediate vicinity of the pore, at
least in our simulations. This is shown in Fig. \ref{fig4}, where for
a tethered polymer of length $N=100$, the average distance $\langle
Z^{(4)}(t=0)\rangle$ of the centre-of-mass of the first $4$ monomers
along the polymer's backbone from the membrane, counting from the
tethered end of a polymer, is plotted versus the chain tension $\Phi$,
for a variety of $h$ values. Within the error bars, all the points in
Fig. \ref{fig4} fall on a straight line, implying that $\Phi$ is very
well-proxied by $\langle Z^{(4)}\rangle$. Since measurements of the
chain tension via Eq. (\ref{e3}) are much more noisy than measurements
of $\langle Z^{(4)}\rangle$, we will use the latter quantity as a
measure for the chain tension.

\section{Memory effects in the $z$-component of the chain tension
at the pore\label{sec3}}

In the case of unbiased polymer translocation, we have witnessed in
Refs. \cite{anom,anomlong} that the memory effects of the polymer give
rise to anomalous dynamics of translocation. We argued
\cite{anom,anomlong} that the velocity of translocation
$v(t)=\dot{s}(t)$, representing monomer current, responds to
$\phi(t)$, the imbalance in the monomeric chemical potential across
the pore acting as ``voltage''. Simultaneously, $\phi(t)$ also adjusts
in response to $v(t)$. In the presence of memory effects, they are
related to each other by
$\phi(t)=\phi_{t=0}+\int_{0}^{t}dt'\mu(t-t')v(t')$ via the memory
kernel $\mu(t)$, which can be thought of as the (time-dependent)
``impedance'' of the system.

In this section, following Refs. \cite{anom,anomlong} we determine the
memory kernel $\mu(t)$ to describe the dynamics of translocation out
of (strong) planar confinements. Note that for the case $h_1=h$, there
is an obvious symmetry between the polymer segments confined within
the parallel plates below and above the $z=0$ plane, implying that the
corresponding memory kernels denoted by $\mu^{(h)}_{z>0}(t)$ and
$\mu^{(h)}_{z<0}(t)$ are the same.  For the case $h_1=\infty$, we
already know the form of $\mu^{(\infty)}_{z>0}(t)$ from
Ref. \cite{anom,anomlong}. In fact, in Ref. \cite{anom} we determined
$\mu^{(\infty)}_{z>0}(t)$ by injecting $p$ monomers into the tethered
end of an equilibrated, tethered polymer of length $N/2-p$ (bringing
the total length to $N/2$), and proxying $\phi(t)$ by the average
distance of the centre-of-mass of the first $4$ monomers $\langle
Z^{(4)}(t)\rangle$ from the membrane. We found
\begin{eqnarray}
\mu^{(\infty)}_{z>0}(t)\sim
t^{-\frac{1+\nu_{\tinytext{3D}}}{1+2\nu_{\tinytext{3D}}}}\exp[-t/\tau_{\text{Rouse(3D)}}]\,,
\label{e4}
\end{eqnarray}
with $\tau_{\text{Rouse(3D)}}$ is the Rouse time for a polymer of
length $N/2$, i.e., $\tau_{\text{Rouse(3D)}}\sim
(N/2)^{1+2\nu_{\tinytext{3D}}}$.

Following Refs. \cite{anom,anomlong}, here we compute
$\mu^{(h)}_{z<0}(t)$, the memory effect of a polymer of length $N/2$
with one end tethered to the pore in the membrane placed at
$z=0$. Indeed, the expression for $\mu^{(h)}_{z<0}(t)$, as we derive
below, is given by
\begin{eqnarray}
\mu^{(\infty)}_{z<0}(t)\sim
t^{-\frac{1+\nu_{\tinytext{2D}}}{1+2\nu_{\tinytext{2D}}}}\exp[-t/\tau_{\text{Rouse(2D)}}]\,.
\label{e5}
\end{eqnarray}
\begin{figure}[!h]
\begin{center}
\includegraphics[width=0.45\linewidth,angle=270]{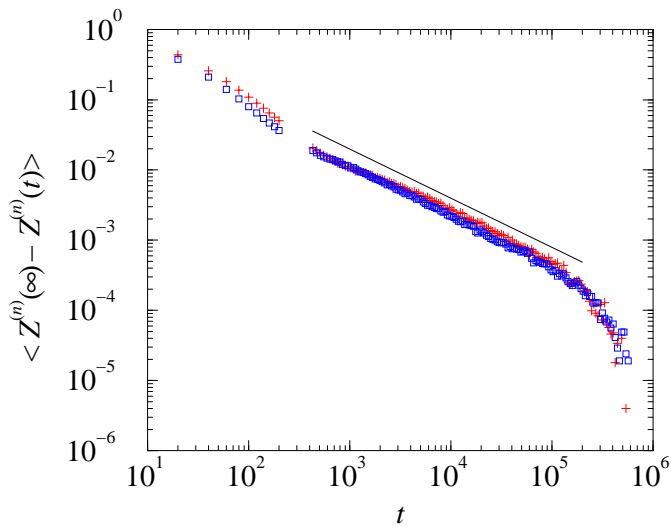}
\caption{$\langle Z^{(4)}(\infty)-Z^{(4)}(t)\rangle$ for $N/2=250$,
for $h=3$ (pluses) and $h=2$ (squares). The solid line corresponds to the
power law
$t^{-(1+\nu_{\tinytext{2D}})/(1+2\nu_{\tinytext{2D}})}$. Note that the
data for $h=2$ obeys a cleaner power law
$t^{-(1+\nu_{\tinytext{2D}})/(1+2\nu_{\tinytext{2D}})}$ than the data
for $h=3$: this we can expect, as the polymer resembles a
two-dimensional polymer more for $h=2$ than for $h=3$. The angular
brackets correspond to $3,200,000$ polymer realisations. \label{fig5}}
\end{center}
\end{figure}

While the Rouse relaxation $\exp[-t/\tau_{\text{Rouse(2D)}}]$ can be
easily justified based on Fig. \ref{fig3}, the value of $\alpha$ for
$\mu^{(\infty)}_{z<0}(t)\sim
t^{-\alpha}\exp[-t/\tau_{\text{Rouse(2D)}}]$ is obtained by following
the procedure of Refs. \cite{anom,anomlong}. The value of $\alpha$
depends on the relaxation properties following the event of injecting,
say, $p$ extra monomers at the tether end, just like extra monomers
add to (or get taken out of) the polymer segment confined within $z<0$
during translocation. Given the $\exp[-t/\tau_{\text{Rouse(2D)}}]$
behaviour of Fig. \ref{fig3}, we anticipate that by time $t$ after the
extra monomers are injected at the tethered point, the extra monomers
will come to a steady state across the inner part of the polymer up to
$n_{t}\sim t^{1/(1+2\nu_{\tinytext{2D}})}$ monomers from the tethered
point, but not significantly further. This internally equilibrated
section of $n_{t}+p$ monomers extends only $r(n_{t})\sim
n_{t}^{\nu_{\tinytext{2D}}}$, less than its equilibrated value
$\left(n_{t}+p\right)^{\nu_{\tinytext{2D}}}$, because the larger scale
conformation has yet to adjust: the corresponding compressive force
from these $n_{t}+p$ monomers is expected by standard polymer scaling
\cite{degennes} to follow $f/(k_{B}T)\sim\delta
r(n_{t})/r^{2}(n_{t})\sim\nu_{\tinytext{2D}}
p/\left[n_{t}r(n_{t})\right]\sim
t^{-(1+\nu_{\tinytext{2D}})/(1+2\nu_{\tinytext{2D}})}$, for $p \ll
n_t$. As was the case in Refs. \cite{anom,anomlong,forced}, we expect
that the chain tension at the pore behaves linearly with the force
$f$, leading to $\alpha=(1+\nu_{\text{2D}})/(1+2\nu_{\text{2D}})=0.7$.

We have confirmed this picture by measuring the impedance response
through simulations. In Fig. \ref{fig4}, we have shown that the
centre-of-mass of the first few monomers is an excellent proxy for
chain tension at the pore and we assume here that this further serves
as a proxy for $\delta\Phi$. Based on this idea, we track
$\langle\delta\Phi^{(z<0)}(t)\rangle$ by measuring $\langle
Z^{(4)}(t)\rangle$, in response to the injection of extra monomers
near the pore at time $t=0$. Specifically we consider the equilibrated
segment of the polymer confined within $z<0$, of length $N/2-5$ (with
one end tethered at the pore), adding $5$ extra monomers at the
tethered end of the polymer segment at time $t=0$, corresponding to
$p=5$, bringing its length up to $N/2$. Using the proxy $\langle
Z^{(4)}(t)\rangle$, we then track
$\langle\delta\Phi^{(z<0)}(t)\rangle$. At short times, these $p$
monomers quickly expand in the $z$-direction (e.g., $t\lesssim400$ in
Fig. \ref{fig5}). Only after they ``feel'' the impenetrable membrane
at $z=-h$, they turn around to expand along the $xy$-plane: indeed,
the agreement between the latter and the theoretical prediction of
$\alpha=(1+\nu_{\tinytext{2D}})/(1+2\nu_{\tinytext{2D}})$ for
$t\gtrsim400$, for $N/2=250$, and for $h=2$ and $3$, can be seen in
Fig. \ref{fig5}. Note that the sharp deviation of the data from the
power law $t^{-(1+\nu)/(1+2\nu)}$ at long times is due to the
asymptotic exponential decay as $\exp[-t/\tau_{\text{Rouse(2D)}}]$ of
$\langle\delta\Phi^{(z<0)}(t)\rangle$ at long times. See also Fig. 2
of Ref. \cite{anom} in this context.

\section{The case $h_1=h$: consequence of the polymer's memory effects
on unbiased translocation\label{sec4}}

For the case $h_1=h$ the condition
$\mu^{(h)}_{z>0}(t)=\mu^{(h)}_{z<0}(t)\equiv\mu^{(h)}(t)$ for the
polymer segments above and below the plane $z=0$ allows us to write
$\phi(t)=\phi_{t=0}+\int_{0}^{t}dt'\mu^{(h)}(t-t')v(t')$, as presented
in the starting paragraph of Sec. \ref{sec3}.   Supposing a
zero-current equilibrium condition at time $t=0$, this relation can be
inverted to obtain $v(t)=\int_{0}^{t}dt'a(t-t')\phi(t')$, where $a(t)$
can be thought of as the ``admittance''. In the Laplace transform
language, $\tilde{\mu}^{(h)}(k)=\tilde{a}^{-1}(k)$, where $k$ is the
Laplace variable representing inverse time. Via the
fluctuation-dissipation theorem, they are related to the respective
autocorrelation functions as
$\mu^{(h)}(t-t')=\langle\phi(t)\phi(t')\rangle_{v=0}$ and
$a(t-t')=\langle v(t)v(t')\rangle_{\phi=0}$.

As explained in the introduction, translocation for $h_1=h$ is
unbiased, for which, having shown that $\mu^{(h)}(t)\sim
t^{-\frac{1+\nu_{\tinytext{2D}}}{1+2\nu_{\tinytext{2D}}}}\exp[-t/\tau_{\text{Rouse(2D)}}]$,
we expect \cite{anom,anomlong} that the translocation dynamics is
anomalous for $t<\tau_{\text{Rouse(2D)}}$, in the sense that the
mean-square displacement of the monomers through the pore,
$\langle\Delta s^{2}(t)\rangle\sim t^{\beta}$ for some $\beta<1$ and
time $t<\tau_{\text{Rouse(2D)}}$, whilst beyond the Rouse time it
becomes simply diffusive. Strictly speaking, $\tau_{\text{Rouse(2D)}}$
in this expression should be replaced by the characteristic
equilibration time of a tethered polymer with length of $O(N)$; since
both scale as $N^{1+2\nu_{\tinytext{2D}}}$, we use
$\tau_{\text{Rouse(2D)}}$ here, favouring notational simplicity. The
value
$\beta=\alpha=\frac{1+\nu_{\tinytext{2D}}}{1+2\nu_{\tinytext{2D}}}$
follows trivially by expressing $\langle\Delta s^{2}(t)\rangle$ in
terms of (translocative) velocity correlations $\left\langle
v(t)v(t')\right\rangle $, which (by the Fluctuation Dissipation
theorem) are given in terms of the time dependent admittance
$a(t-t')$, and hence inversely in terms of the corresponding
impedance. In other words, up to the Rouse time, the squared
displacement as a function of time is subdiffusive, following
$\langle\Delta s^{2}(t)\rangle\sim
t^{\frac{1+\nu_{\tinytext{2D}}}{1+2\nu_{\tinytext{2D}}}}$. Consequently,
at the Rouse time $\tau_{\text{Rouse(2D)}}\sim
N^{1+2\nu_{\tinytext{2D}}}$, the squared displacement scales as
$\langle\Delta s^{2}[\tau_{\text{Rouse(2D)}}]\rangle\sim
N^{1+\nu_{\tinytext{2D}}}$.  Beyond the Rouse time, there are no
memory effects and the squared displacement increases linearly in
time: $\langle\Delta s^{2}(t)\rangle \sim\{\langle\Delta
s^{2}[\tau_{\text{Rouse(2D)}}]\rangle/\tau_{\text{Rouse(2D)}}\} t\sim
N^{-\nu_{\tinytext{2D}}} t$. Based on the criterion for unthreading,
i.e., unthreading occurs when $\sqrt{\langle\Delta
s^{2}(\tau_{d})\rangle}\sim N$, one then obtains $\tau_d \sim
N^{2+\nu_{\tinytext{2D}}}$ \cite{anom,anomlong}.
\begin{table}[!h]
\begin{center}
\begin{tabular}{c|c|c|c}
$\quad\quad N\quad\quad $ & $\quad\quad \tau_u\quad\quad$ & $\quad
\tau_u/N^{2+\nu_{\tinytext{2D}}}\quad$ & $\quad
\tau_u/N^{1+2\nu_{\tinytext{2D}}}\quad $ \tabularnewline \hline \hline
30& 2439 & 0.2114 & 0.4948 \tabularnewline \hline 40 & 5176 & 0.2034 &
0.5115 \tabularnewline \hline 50 & 9499 & 0.2021 & 0.5373
\tabularnewline \hline 60 & 15684 & 0.2021 & 0.5624 \tabularnewline
\hline 70 & 24532 & 0.2069 & 0.5984\tabularnewline \hline 80 & 34556 &
0.2018 & 0.6037\tabularnewline \hline 90 & 47974 & 0.2027 &
0.6243\tabularnewline \hline 100 & 64755 & 0.2048 &
0.6476\tabularnewline \hline 200 & 415767 & 0.1954 &
0.7350\tabularnewline \hline 300 & 1268463 & 0.1955 &
0.8137\tabularnewline \hline 400 & 2765246 & 0.1932 &
0.8641\tabularnewline \hline 500 & 4961331 & 0.1877 &
0.8875\tabularnewline \hline 600 & 8228721 & 0.1885 &
0.9332\tabularnewline \hline 700 & 12648891 & 0.1897 &
0.9758\tabularnewline \hline 800 & 17975330 & 0.1867 &
0.9930\tabularnewline \hline
\end{tabular}
\caption{Median values of $\tau_u$ based on $8,192$ unthreading
realisations for each $N$.\label{table2}}
\end{center}
\end{table}

\begin{figure}[t]
\begin{center}
\includegraphics[width=0.55\linewidth]{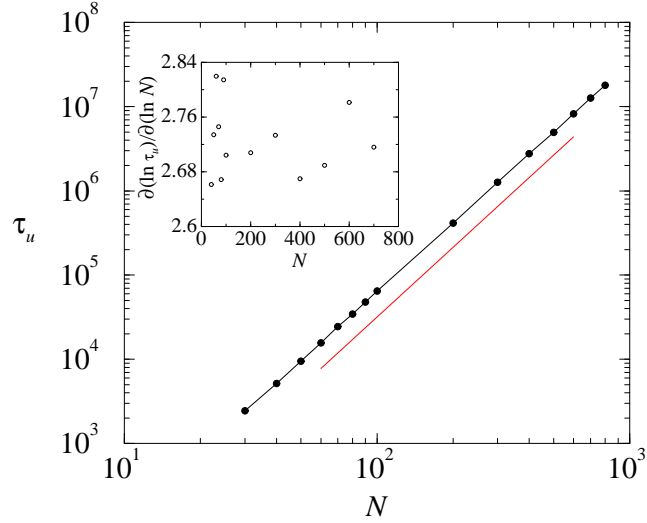}
\caption{Scaling of $\tau_u$ with $N$: $\tau_u$ data of Table
\ref{table2} are represented by the black line with points, solid line
corresponds to the scaling $\tau_u\sim N^{2+\nu_{\tinytext{2D}}}$. The
effective exponents $\partial(\ln\tau_u)/\partial(\ln N)$ for
$N=40,50,60,70,80,90,100,200,300,400,500,600,700$ are shown in the
inset, clearly ruling out $N^{1+2\nu_{\tinytext{2D}}}$ scaling of
$\tau_u$ in favour of $N^{2+\nu_{\tinytext{2D}}}$.\label{fig6}}
\end{center}
\end{figure}

\begin{figure}[t]
\begin{center}
\includegraphics[width=0.6\linewidth]{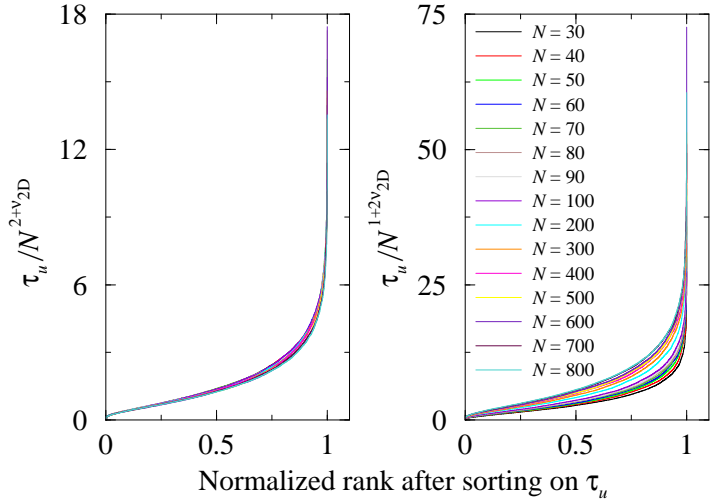}
\caption{Sorted values $\tau_u/N^{2+\nu_{\tinytext{2D}}}$ and
$\tau_u/N^{1+2\nu_{\tinytext{2D}}}$ vs. their normalised rank, for
$8,192$ realisations per value of $N$. Data shown (from bottom to top
in the right panel):
$N=30,40,50,60,70,80,90,100,200,300,400,500,600,700,800$. The left
panel shows a much better collapse than the right panel, ruling out
the $\tau_d$ scaling as $N^{1+2\nu_{2D}}$, as observed by Luo et
al. \cite{luo1}. In fact, the data collapse in the left panel suggests
that $P(\tau_d)$, the probability distribution for the dwell time
$\tau_d$ has a scaling form $P(\tau_d)\sim{\cal
P}(\tau_d/N^{2+\nu_{\tinytext{2D}}})/N^{2+\nu_{\tinytext{2D}}}$, with
a scaling function ${\cal P}(x)$. \label{fig7}}
\end{center}
\end{figure}

For computer simulations of unbiased translocation in two dimensions,
Luo et al. \cite{luo1} reported a scaling of $\tau_{d}\sim
N^{1+2\nu_{\tinytext{2D}}}$; note that this exponent and our
theoretical expectation is $10\%$ different.  To distinguish these two
different exponents, we performed high-precision simulations to obtain
the unthreading time for a number of $N$-values:
$N=30,40,50,60,70,80,90,100,200,300,400,500,600,700,800$, for $8,192$
realisations for each value of $N$, with a membrane spacing of $h=3$
(Table \ref{table2}). The unthreading time $\tau_u$ in Table
\ref{table2}) is defined as the time for the polymer to leave the pore
with $s(t=0)=N/2$ and the two polymer segments equilibrated at
$t=0$. Both $\tau_u$ and $\tau_d$ scale the same way, since
$\tau_u<\tau_d<2\tau_u$ \cite{anomlong}. The data of Table
\ref{table2}, along with the effective exponent for $\tau_u$ vs. $N$
as a power law are further shown in Fig. \ref{fig6}. To be able to use
the full potential of the statistics of $8,192$ realisations for each
$N$, we also plot the sorted $\tau_u/N^{2+\nu_{\tinytext{2D}}}$ and
$\tau_u/N^{1+2\nu_{\tinytext{2D}}}$ vs. the normalised rank of the
sorted values in Fig. \ref{fig7}. The data collapse is a further test
that rules out the $\tau_d$ scaling as $N^{1+2\nu_{2D}}$, as reported
in Ref. \cite{kantor1,luo1,wei}: in fact the left panel of Fig. 
\ref{fig7} suggests that $P(\tau_d)$, the probability distribution for 
the dwell time $\tau_d$ has a scaling form $P(\tau_d)\sim{\cal
P}(\tau_d/N^{2+\nu_{\tinytext{2D}}})/N^{2+\nu_{\tinytext{2D}}}$, with
a scaling function ${\cal P}(x)$. These results together clearly
demonstrate that the dwell time scales as $\tau_d \sim
N^{2+\nu_{\tinytext{2D}}}$.
\clearpage

\section{$\tau_d$ for $h_1=\infty$, or equivalently, $\tau_d$ for
  field-driven translocation in two dimensions\label{sec5}}

\subsection{Lower bound for $\tau_d$ for field-driven
  translocation\label{sec5a}}

We now turn to the case $h_1=\infty$, which is equivalent to
field-driven translocation in two dimensions as we discussed in the
introduction. For our first stop, we notice in Table \ref{table0b}
that quite a few reported numerical results violate the lower bound
$N^{1+\nu}$ for the scaling exponent of $\tau_d$, suggested in
Ref. \cite{kantor2}, including that of the authors of
Ref. \cite{kantor2} themselves. In light of this, below we first
discuss the lower bound for field-driven translocation.

The crux of the derivation of the lower bound for $\tau_d$ in
Ref. \cite{kantor2} is that, with or without an applied field, the
mobility of a polymer translocating through a narrow pore in a
membrane will not exceed that of a polymer in bulk (i.e., in the
absence of the membrane). To obtain the mobility of a polymer in bulk,
the authors assumed two more attributes of a polymer under a driving
field:
\begin{itemize}
\item[(i)] To mimic the action of a field on a translocating polymer,
the field on the polymer in bulk has to act on a monomer whose
position along the backbone of the polymer changes continuously in
time. As a result, there is no incentive for the polymer to change its
shape from its bulk equilibrium shape, i.e., the polymer can still be
described by a blob with radius of gyration $\sim N^\nu$ in the
appropriate dimension.
\item[(ii)] The polymer's velocity is $\sim mE$, where $E$ is the
field, and $m$ is the mobility $\sim1/N$.
\end{itemize}
Of these two assumptions, note that (ii) is obtained as the steady
state solution of the equation of motion of a Rouse polymer, in bulk,
with uniform velocity and vanishing internal forces (see e.g.,
Ref. \cite{degennes}, Eq. VI.10). We have already witnessed in many
occasions
\cite{kantor1,kantor2,klafter,dubbeldam,dubbeldam2,anom,anomlong,forced}
that the dynamics of translocation through a narrow pore is anomalous
(subdiffusive), and in Sec. \ref{sec3} of this paper we have seen that
there are strong memory effects in the polymer, to the point that the
velocity of translocation is not constant in time. The anomalous
dynamics and the memory effects are crucial ingredients that question
the validity of the lower bound $N^{1+\nu}$ for $\tau_d$ for
field-driven translocation.

A lower bound for $\tau_d$ for field-driven translocation does
nevertheless exist, and it can be obtained from conservation of
energy. Consider a translocating polymer under an applied field $E$
which we can assume to be acting only at the pore: $N$ monomers take
time $\tau_d$ to translocate through the pore. The total work done by
the field in time $\tau_d$ is then given by $EN$. In time $\tau_d$,
each monomer travels a distance of $\sim R_g$, leading to an {\it
average\/} monomer velocity $v_m\sim R_g/\tau_d$. The rate of loss of
energy due to viscosity $\eta$ of the surrounding medium per monomer
is given by $\eta v_m^2$. For a Rouse polymer, the frictional force on
the entire polymer is a sum of frictional forces on individual
monomers, leading to the total free energy loss due to the viscosity
of the surrounding medium during the entire translocation event
scaling as $\Delta F\sim N\tau_d\eta v_m^2=N\eta R^2_g/\tau_d$. This
loss of energy must be less than or equal to the total work done by
the field $EN$, which yields us the inequality $\tau_d\geq\eta
R^2_g/E=\eta N^{2\nu}/E$ \cite{noteunbiased}.

\subsection{$\tau_d$ for the case $h_1=\infty$\label{sec5b}}

If we follow the procedure due to two of us in Ref. \cite{forced} to
calculate $\tau_d$ for the case $h_1=\infty$ via the memory kernels
discussed in Sec. \ref{sec3}, then we would adopt the following
route. For the translocated part of the polymer (in the space $z>0$),
the memory kernel takes the form of Eq. (\ref{e4}), while the
translocating part of the polymer (in the space $z<0$ and $z>-h$) the
memory kernel takes the form of Eq. (\ref{e5}). Of these, the
magnitude of the exponent in the power law of Eq. (\ref{e5}) is less
than that of Eq. (\ref{e4}) [i.e., the memory effects of the
translocating part of the polymer are longer-lived than that of the
translocated part of the polymer], which implies that the relation
between the chain tension imbalance across the pore and translocation
velocity should be described by the equation
\begin{eqnarray}
\phi(t)=\phi_{t=0}-\int_0^t dt'\,|\mu^{(\infty)}_{z<0}(t-t')|\,v(t')\,,
\label{e6}
\end{eqnarray}
leading to
\begin{eqnarray}
v(t)=\int_0^t
dt'\,(t-t')^{-(1+3\nu_{\tinytext{2D}})/(1+2\nu_{\tinytext{2D}})}\,[\phi(0)-\phi(t')]\,,
\label{e7}
\end{eqnarray}
via Laplace transform \cite{forced}. Furthermore, if
$[\phi(0)-\phi(t')]$ remains a constant (not shown here
\cite{notetension}), then Eq. (\ref{e7}) yields $v(t)\sim
t^{-\nu_{\tinytext{2D}}/(1+2\nu_{\tinytext{2D}})}$, implying that the
distance $[s(t)-s(0)]$ unthreaded in time $t$ should behave as
$s(t)=s(0)+\displaystyle{\int^t_0dt'\,v(t')\sim
t^{(1+\nu_{\tinytext{2D}})/(1+2\nu_{\tinytext{2D}})}}$. With
$[s(\tau_d)-s(0)]=N$, the relation $[s(t)-s(0)]\sim
t^{(1+\nu_{\tinytext{2D}})/(1+2\nu_{\tinytext{2D}})}$ would finally
yield $\tau_d\sim
N^{(1+2\nu_{\tinytext{2D}})/(1+\nu_{\tinytext{2D}})}$.

The scaling $\tau_d\sim
N^{(1+2\nu_{\tinytext{2D}})/(1+\nu_{\tinytext{2D}})}$ obtained through
the memory kernel approach violates the lower bound $\tau_d\sim
N^{2\nu_{\tinytext{2D}}}$, and therefore the former cannot be the
correct scaling for $\tau_d$. Indeed a short reflection makes the
issue clear. The memory kernels we derived in Sec. \ref{sec3} are
actually ``static memory kernels'': the individual monomer velocities
involved, for determining the static memory kernels (i.e., for the
equilibration process of the polymer when $p$ extra monomers are
injected at the pore, with $p\ll N$), are small, and as a result, the
energy loss due to viscosity of the surrounding medium is
negligible. For field-driven translocation in two-dimensions, since
the lower bound for $\tau_d$ set by the viscous energy loss overrides
the expression of $\tau_d$ obtained from the static memory kernel, it
is very well possible that there exists a corresponding ``{\it
dynamic\/} memory kernel'' that describes the relations between
$\phi(t)$ and $v(t)$ for a translocating polymer.

\begin{figure}[!t]
\begin{center}
\includegraphics[width=0.5\linewidth,angle=270]{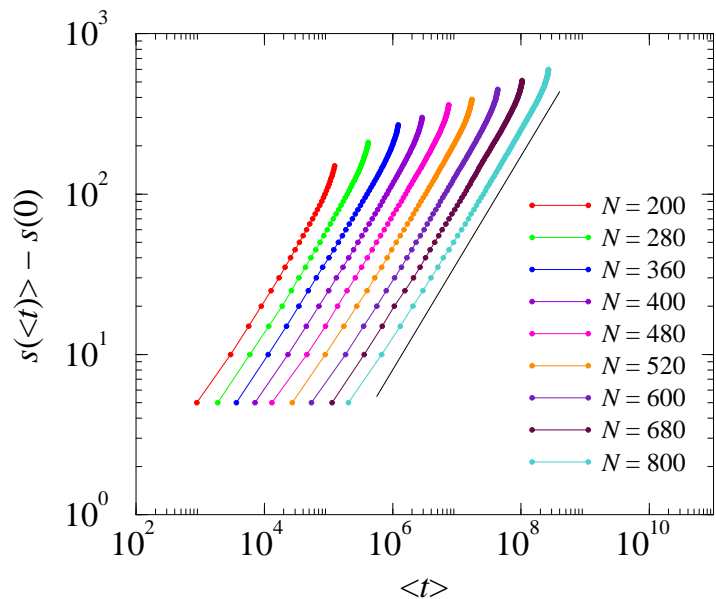}
\caption{The mean time $\langle t\rangle$ required to unthread a
distance $[s-s(0)]$ for $h=3$, $h_1=\infty$, $s(0)=3N/4$ and
$[s-s(0)]=5,10,\ldots,N$. The $\langle t\rangle$ values are obtained
as average over $2,048$ realisations for each $N$. The $N$-values
used are (from left to right):
$200,280,360,400,480,520,600,680,800$. The $\langle t\rangle$-values
for $N=200$ is the actual value of time, for the others, $\langle
t\rangle$-values for each $N$ are separated by a factor $2$ along the
$x$-axis. The solid line represents a power law with exponent
$1/(2\nu_{\tinytext{2D}})$, corresponding to $\tau_d\sim
N^{2\nu_{\tinytext{2D}}}$.\label{fig8}}
\end{center}
\end{figure}

We have not found a way to probe this dynamic memory kernel. We can
nevertheless view its consequences by plotting the distance unthreaded
$[s(t)-s(0)]$ as a function of $t$. We chose $h=3$ for a variety of
lengths of polymers (corresponding to strong confinement, i.e.,
equivalent to translocation in two dimensions driven by a strong
field) for this purpose. For a polymer of length $N$ we hold $3N/4$
monomers between the plates $z=0$ and $z=-h$, with the other $N/4$
monomers protruding out through the pore in the space $z>0$, i.e.,
$s(0)=3N/4$ \cite{notematter}. We equilibrate both segments, and at
$t=0$ we let translocation begin. The mean time $\langle t\rangle$
required to unthread a distance $[s-s(0)]$ and the scaling of $\tau_d$
are presented in Fig. \ref{fig8} and Table \ref{table3}
respectively. We note that although $N^{2\nu_{\tinytext{2D}}}$
provides only the lower bound for $\tau_d$, our simulation data, along
with those of Refs. \cite{kantor2,luijten} suggests that
$N^{2\nu_{\tinytext{2D}}}$ is indeed the correct scaling for the
$\tau_d$ in two dimensions.
\begin{table}[!t]
\begin{center}
\begin{tabular}{c|c|c}
$\quad\quad N\quad\quad $ & $\quad\quad \tau_d\quad\quad$ & $\quad
\tau_u/N^{2\nu_{\tinytext{2D}}}\quad$ \tabularnewline \hline \hline
200& 117934 & 41.6960 \tabularnewline \hline  280 & 198497 & 42.3659
\tabularnewline \hline  360 & 294199 & 43.0712 \tabularnewline \hline
400 & 358865 & 44.8581 \tabularnewline \hline  480 & 454131 & 43.1836
\tabularnewline \hline  520 & 519663 & 43.8245 \tabularnewline \hline
600 & 653885 & 44.4912 \tabularnewline \hline  680 & 779484 & 43.9586
\tabularnewline \hline  800 & 1026083 & 45.3469 \tabularnewline \hline
\end{tabular}
\caption{Median values of $\tau_d$ for $h=3$, demonstrating
$\tau_d\sim N^{2\nu_{\tinytext{2D}}}$. The values are based on $2,048$
translocation realisations for each $N$.\label{table3}}
\end{center}
\end{table}

\section{Discussion\label{sec6}}

Polymer translocation out of confined spaces plays an important role
for polymers in various biological processes. Planar confinements,
however, are mostly a theoretical construct. In this paper, we have
studied polymer translocation in three dimensions out of planar
confinements, and demonstrated that polymer translocation in these
(theoretically motivated) geometries is a very interesting testcase of
fundamental physics of translocation dynamics.

The geometry we have considered is as follows. We have divided the
three-dimensional space into two parts: $z>0$ and $z<0$ by a membrane
placed at $z=0$. This membrane is impenetrable to the polymer except
for a narrow pore. We have then placed two more parallel membranes at
$z=-h$ and $z=h_1$ that are completely impenetrable to the
polymer. The polymer is initially sandwiched between the membranes
placed at $z=-h$ and $z=0$. We have considered strong confinement for
the polymer, i.e., $h\ll R^{(\tinytext{3D})}_g$ where
$R^{(\tinytext{3D})}_g$ is its radius of gyration in the bulk scaling
as $\sim N^{\nu_{\tinytext{3D}}}$. Here, $N$ is the polymer length and
$\nu_{\text{3D}}\simeq0.588$ is the Flory exponent in three
dimensions. Under these conditions the confined segment of the polymer
essentially behaves as a polymer in two dimensions. If $h_1>h$, the
initial state of the polymer is entropically unfavourable, and the
polymer escapes to the space between the membranes placed at $z=0$ and
$z=h_1$ through the pore in the membrane placed at $z=0$: this is
essentially the field-driven translocation process in two
dimensions. We have studied two separate cases: (i) $h_1=h$, and (ii)
$h_1=\infty$. For (i) the entropic drive is absent; the translocation
dynamics reduces to that of an unbiased translocation for a
two-dimensional polymer. Based on theoretical analysis and
high-precision simulation data, we have shown that the dwell time
$\tau_d$, the time the pore remains occupied during translocation,
scales as $N^{2+\nu_{\tinytext{2D}}}$, in perfect agreement with our
previous results. We have also shown that the probability distribution
of the dwell time, $P(\tau_d)$ for $h_1=h$ has a scaling form
$P(\tau_d)\sim {\cal
P}(\tau_d/N^{2+\nu_{\tinytext{2D}}})/N^{2+\nu_{\tinytext{2D}}}$. For
(ii) we have shown that $\tau_d\sim N^{2\nu_{\tinytext{2D}}}$, in
agreement with several existing numerical results in the
literature. Here $\nu_{\text{2D}}=0.75$ is the Flory exponent in two
dimensions. The result $\tau_d\sim N^{2\nu_{\tinytext{2D}}}$ for case
(ii) violates the earlier reported lower bound $1+N^\nu$ for $\tau_d$
for field-driven translocation. We have argued, based on conservation
of energy, that the actual lower bound for $\tau_d$ is $N^{2\nu}$ and
not $1+N^\nu$.

\section*{Acknowledgements}

We gratefully acknowledge ample CPU time at the Dutch national
supercomputer cluster SARA.


\begin{thebibliography}{07}
\bibitem{noteunbiased} For unbiased translocation, $EN$ in this
argument is to be replaced by the difference in free energy (or
entropy) of a threaded polymer, corresponding to $s=N/2$, and the
translocated polymer, corresponding to $s=N$. This leads to the
inequality $\tau_d\geq\eta N^{1+2\nu}$.
\bibitem{notetension} We have checked, for $h=3$, using the proxy
variable $Z^{(4)}$ for $\Phi$ as in Fig. 3(a) of Ref. \cite{forced},
that $[\phi(0)-\phi(t)]$ indeed remains a constant throughout
translocation, aside from some small values of $t$.
\bibitem{notematter} We have checked that the scaling results
are not influenced by whether $s(0)=N/4$ or $N/2$.
\end{thebibliography}
\end{document}